\documentstyle[sprocl,psfig]{article}

\input{psfig}

\def\Journal#1#2#3#4{{#1} {\bf #2}, #3 (#4)}

\def\PLB{{\it Phys. Lett.}  B}
\def\PRL{\it Phys. Rev. Lett.}
\def\PRD{{\it Phys. Rev.} D}

\def\EPJ{{\it Eur. Phys. J.} C}

\def\d0{D^0}
\def\d0b{\bar{D^0}}

\def\be{\begin{equation}}
\def\ee{\end{equation}}
\def\bea{\begin{eqnarray}}
\def\eea{\end{eqnarray}}
\begin{document}

\title{CHARM-SYSTEM TESTS OF CPT WITH FOCUS\footnote{Invited Talk
given at the Second Meeting on CPT and Lorentz Symmetry, 
Indiana University, Bloomington, August 15-18, 2001.}}

\author{R.W. GARDNER}

\address{Department of Physics, Indiana University \\
701 East Third Street, 
Bloomington, IN 47405, USA\\E-mail: rwg@indiana.edu}


\maketitle\abstracts{ 
We discuss a search for CPT violation in neutral charm meson
oscillations.  The data come from the Fermilab fixed-target 
experiment FOCUS.  
While flavor mixing in the charm sector is predicted to be 
small by the standard model, it is still possible to 
investigate CPT violation through study
of the proper time dependence of the asymmetry in
right-sign decay rates for $D^0$ and $\bar{D^0}$.
Using present limits for $D^0-\bar{D^0}$ mixing
we infer bounds on charm CPT violation using data 
from FOCUS.
}

\section{Introduction}

The combined symmetry of charge conjugation (C), parity (P), and 
time reversal (T) is believed to be respected
by all local, point-like, Lorentz covariant field 
theories, such as the standard model.  
However, extensions to the standard model based on string 
theories do not necessarily require CPT invariance,
and observable effects at low-energies may be within in
reach of experiments studying flavor 
oscillations.\cite{kostelecky-95}
A framework\cite{kostelecky} in which indirect
CPT and T violating parameters
appear has been developed which allows
experimental investigation in neutral meson systems.

For quite some time searches for CPT violation have been
made in the neutral kaon system in which 
particle-antiparticle mixing is large.  
KTeV\cite{ktev} reports a bound on the CPT figure of merit 
$r_K \equiv |m_{K^0} - m_{\bar{K}^0}|/m_{K^0} < (4.5 \pm 3) \times 10^{-19}$.  
CPT tests in $B^0$ meson decay have been 
made by OPAL\cite{opal} at LEP, and by
Belle at KEK which has recently 
reported\cite{belle} 
$r_B \equiv |m_{B^0} - m_{\bar{B}^0}|/m_{B^0} < 1.6 \times 10^{-14}$.

To date, no experimental search for CPT violation has been
made in the charm quark sector.  This is probably due in part to 
the expected suppression of $D^0-\bar{D}^0$  oscillations
in the standard model, and the lack of a strong mixing
signal in the experimental data.  Recent mixing searches include
a study of lifetime differences between charge-parity (CP) eigenstates
from FOCUS, which  reported\cite{focusycp} a value for 
the parameter $y_{CP} = (3.42 \pm 1.39 \pm 0.74)\%$.  
The CLEO Collaboration has reported 
95\% confidence level bounds  on mixing parameters $x'$ and $y'$ 
(related to the usual parameters $x$ and $y$ by a strong phase 
shift):\cite{cleo}
$(1/2)x'^2 < 0.041\%$ and $-5.8\% < y' < 1\%$.  
FOCUS has reported\cite{link} a study of the doubly Cabibbo 
suppressed ratio ($R_{DCS}$)  for the decay $D^0\rightarrow K^+\pi^-$ and has extracted
a contour limit on $y'$ (of order few \%) under varying assumptions of
$R_{DCS}$ and $x'$.
The question
arises -- what can be learned about indirect CPT violation 
given the apparent smallness of mixing in the charm
system? It turns out that even in the absence of a strong mixing 
signal one can still infer the level of CPT violation 
sensitivity through study of the time dependence of 
$D^0$ decays, which we show in this paper.

\section{Proper Time Asymmetry}

Time dependent decay probabilities into right-sign and wrong
sign decay modes for neutral mesons which express the CPT
violation have been developed in a general framework.\cite{kostelecky}
For the decay of $D^0$ to a right-sign final state $f$ (which could
be a semileptonic mode, or a Cabibbo favored hadronic mode), 
the time dependent decay probability is
\begin{eqnarray}
P_f(t) & \equiv & | \langle f | T | D^0(t) \rangle |^2                         \nonumber  \\
       &=& {1\over{2}} |F|^2 {\rm exp}({-{\gamma \over{2}}t})  
           \times [ ( 1 + |\xi|^2 ) {\rm cosh} \Delta\gamma t/2 + 
           (1 - |\xi|^2 ) {\rm cos} \Delta m t                                   \nonumber \\
       &&  - 2~{\rm Re}~\xi~{\rm sinh} \Delta\gamma t/2 
           - 2~{\rm Im}~\xi~{\rm sin}\Delta m t              ].
\label{eq:rsdecay}
\end{eqnarray}

\noindent
The time dependent probability for the  decay of $\bar{D}^0$
to a right-sign final state $\bar f$,  
$\bar{P}_{\bar f}(t)$, may be obtained by replacing in the above equation 
$\xi \rightarrow - \xi$ and
$F \rightarrow \overline{F}$.  
In the formula, $F$ represents the basic
transition amplitude for the decay $D^0\rightarrow f$,
$\Delta\gamma$ and $\Delta m$ are
the differences in physical decay widths and masses for the propagating eigenstates
and can be related to the usual mixing parameters $x  = \Delta M/\Gamma$
and $y = \Delta \Gamma/{2 \Gamma}$.  
The complex parameter $\xi$ controls the CPT violation and is seen to modify
the shape of the time dependent decay probabilities.
Expressions for wrong-sign decay probabilities involve both CPT and T violation
parameters which scale the probabilities, leaving the shape unchanged.
Using only right-sign decay modes, the following asymmetry can be formed,
\begin{eqnarray}
A_{CPT}(t)  =  {{\bar{P}_{\bar f}(t) - P(t)}\over{\bar{P}_{\bar f}(t) + P(t)}},
\label{eq:asym}
\end{eqnarray}

\noindent
which is sensitive to CPT violating parameter $\xi$:
\begin{eqnarray}
A_{CPT} = {{ 2~{\rm Re}~\xi~{\rm sinh}\Delta\gamma t/2 + 2~{\rm Im}~\xi~{\rm sin} \Delta m t} \over {
 ( 1 + |\xi|^2 ) {\rm cosh} \Delta\gamma t/2 + 
           (1 - |\xi|^2 ) {\rm cos} \Delta m t 
}}.
\label{eqn:acpt}
\end{eqnarray}

\begin{figure}[tbh]
\begin{centering}
\psfig{figure=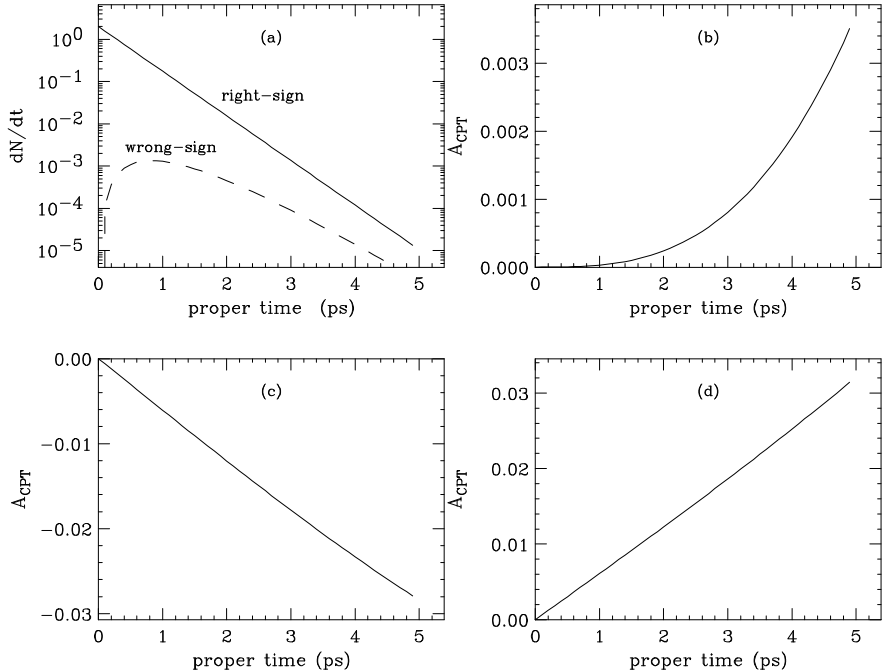,height=3.5in}
\caption{
(a) Proper time decay probabilities for
    $D^0$ decay, right-sign (solid) and wrong-sign (dashed) and
(b) $A_{CPT}$ with  Re~$\xi$ = 5\%, Im~$\xi$ = 5\%.
(c) $A_{CPT}$ with  Re~$\xi$ = 0,   Im~$\xi$ = 5\%.
(d) $A_{CPT}$ with  Re~$\xi$ = 5\%, Im~$\xi$ = 0.
 }  
\label{fig:asym}
\end{centering}
\end{figure}

We can gain insight into the anticipated experimental sensitivity by plotting
these functions with some reasonable assumptions.  We use  95\% confidence level 
upper bounds on the mixing parameters $x$ and  $y$ of 5\%,  which is at the 
upper range of the current experimental sensitivity, as discussed previously.  
In Fig.~\ref{fig:asym}(a) we plot the proper time decay probabilities for
$D^0$ decay under the assumption of CPT violation at the level of 
Re~$\xi$ =5\%, Im~$\xi$ = 5\%, which are independent parameters in the framework.  
One sees a CPT violation-induced wrong-sign contribution which vanishes at
zero proper time and at long proper times.  This causes a distortion
from a purely exponential decay of a $D^0$ (and $\bar{D}^0$), which is then visible in the
asymmetry plot, $A_{CPT}$ as shown in Fig.~\ref{fig:asym}(b).  
Because of the small oscillation frequency
and short lifetime, one sees only the start of the oscillation, growing 
beyond 0.3\% at long proper times.  
Evident from Eqn.~\ref{eqn:acpt} is that positive values of
the Re~$\xi$ and Im~$\xi$  work to oppose one another in the
asymmetry in a linear fashion.  
This is shown in the nearly linear behavior of $A_{CPT}$
in Figs.~\ref{fig:asym}(c,d) which have 
Im~$\xi = 5\%$, Re~$\xi = 0$  and
Im~$\xi = 0$, Re~$\xi = 5\%$ respectively, and consequently CPT
asymmetries larger by a factor of 10 at long proper times .  
In practice, experiments will be sensitive to either Re~$\xi$ or
Im~$\xi$, but not both simultaneously.

\section{Data Analysis}

In this paper we investigate the current
experimental sensitivity for a CPT violating signal
using data  collected by the FOCUS Collaboration during the 
1996-97 fixed-target run at Fermilab.  FOCUS is an
upgraded version of the E687 spectrometer. Charm particles
are produced by the interaction of high energy photons
(average energy $\approx$ 180 GeV for triggered charm states) 
with a segmented
BeO target.  In the target region charged particles are
tracked by 16 layers of microstrip detectors.  These detectors
provide excellent vertex resolution.  Charged particles
are further tracked by a system of five multiwire proportional
chambers and are momentum analyzed by two oppositely 
polarized large aperture dipole magnets.  Particle
identification is determined by three multi-cell 
threshold ${\check{\rm C}}$erenkov detectors, electromagnetic calorimeters,
and muon counters.

The data analysis is as follows.
We analyze the two right-sign hadronic 
decays $D^0 \rightarrow K^-\pi^+ $ 
and $\bar{D}^0 \rightarrow K^+\pi^-$.
Contamination by doubly Cabibbo suppressed (wrong-sign) decay 
occurs but is small owing to the small branching ratio, and
its effect on the right-sign signal yield estimates in this
analysis is negligible.
We use the soft pion from the decay $D^{*+}\rightarrow D^0\pi^+$
to tag the flavor of the $D$ at production, and 
the kaon charge in the decay $D^0\rightarrow K^- \pi^+$
to tag the $D$ flavor at decay. (Charge conjugates are
assumed in this paper.)
$D^0\rightarrow K^- \pi^+$ events were selected by requiring
a minimum detachment of the secondary (decay) vertex from the 
primary (production) vertex of 5 $\sigma_\ell$.  
The primary vertex was found using a candidate driven vertex
finder which nucleated tracks about a ``seed'' track 
constructed using the secondary vertex and the reconstructed
$D$ momentum vector. Both primary and secondary vertices
were required to have confidence level fits of greater
than 1\%.
The $D^*$-tag is accomplished by requiring the
$D^*-D^0$ mass difference to be less than 3 MeV/c$^2$ of
the nominal value.\cite{pdg}

\begin{figure}[tbh]
%\begin{centering}
\centerline{
\psfig{figure=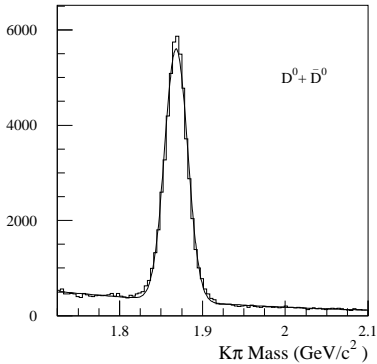,height=2.0in}
}
\caption{Invariant mass distribution for the sum of $D^0$ and 
$\hat{D}^0$ right-sign decay candidates.}  
\label{fig:signal}
%\end{centering}
\end{figure}

Kaons and pions were identified using
the $\check{\rm C}$erenkov particle identification cuts.
These cuts are based on likelihood ratios between the
various stable particle hypotheses, and are computed for a 
given track from the observed firing response (on or off)
of all cells within the track's ($\beta$ = 1) $\check{\rm C}$erenkov 
light cone in each of three multi-cell, threshold $\check{\rm C}$erenkov
counters.    The probability that a given
track wil fire a given cell is computed using
Poisson statistics based on the predicted number of photoelectrons
striking the cell's phototube under each particle 
identification hypothesis; the accidental firing rate
of the cells, which depended on beam intensity, was
also used.  The product of all firing probabilities
for all cells within the three $\check{\rm C}$erenkov cones
produces a $\chi^2$-like variable called
$W_i \equiv $ -- 2$\times$log(likelihood) where $i$ ranges
over electron, pion, kaon and proton hypotheses.
For the $K$ and the $\pi$ candidates we require $W_i$ to be no more
than 4 greater than the smallest of the other three hypotheses
($W_i - W_{min} < 4$) which eliminated candidates that highly
likely to be misidentified. 
In addition, $D^0$ daughters must satisfy the slightly stronger
$K\pi$ separation criteria $W_\pi - W_K > 0.5$ for the $K$
and $W_K - W_\pi > -2 $ for the $\pi$.
Doubly misidentified $D^0 \rightarrow K^-\pi^+$  is removed
by imposing a  hard $\check{\rm C}$erenkov cut on the sum of 
the two separations 
$((W_\pi - W_K)_K + (W_K - W_\pi)_\pi > 8) $ when the 
invariant mass of the track pair with the $K$ and $\pi$
hypothesis exchanged is within 4$\sigma$ of the 
nominal $D^0$ mass.
Fig.~\ref{fig:signal} shows the invariant mass distribution
for two $D^*$-tagged, right-sign decays
$D^0\rightarrow K^-\pi^+$ and $\bar{D}^0\rightarrow K^+\pi^-$.
Approximately 49700 signal events were used in the analysis
described below.

\begin{figure}[tbh]
{\hbox
{\psfig{figure=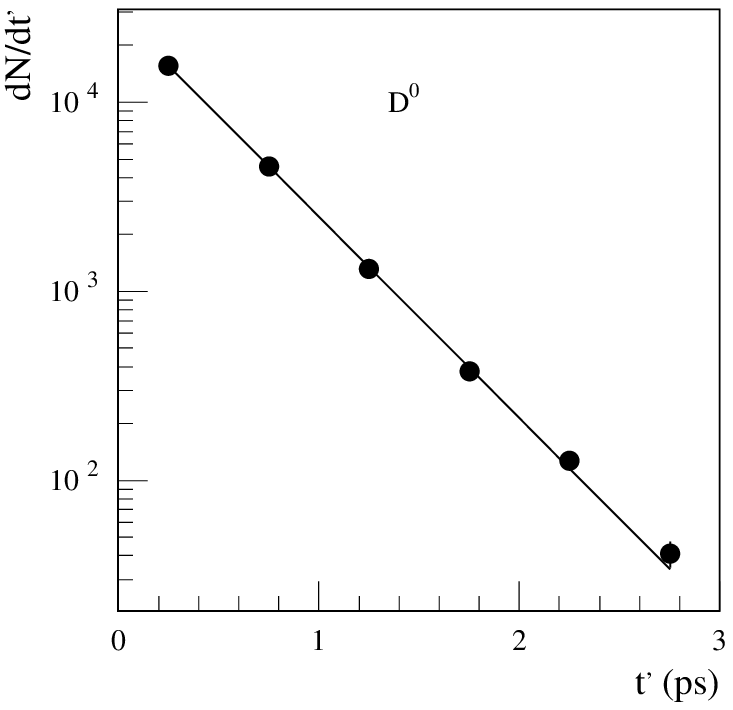,height=2.0in}
\hskip 0.25in
\psfig{figure=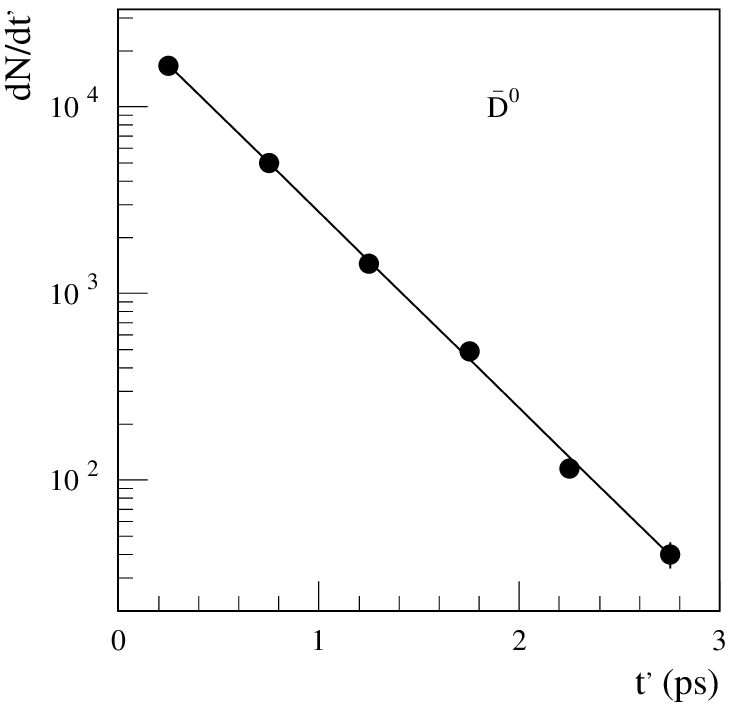,height=2.0in}}
}
\caption{Background
subtracted reduced proper time distributions for 
$D^0$ and $\d0b$. }  
\label{fig:ptime}
\end{figure}

The reduced proper time is a traditional lifetime variable used
in fixed-target experiments which use the detachment between
the primary and secondary vertex as their principal tool in
reducing non-charm background.  The reduced proper time is
defined by $t'=(\ell - N\sigma_\ell)/(\beta\gamma c)$ where
$\ell$ is the distance between the primary and secondary
vertex, $\sigma_\ell$ is the resolution on $\ell$, and $N$
is the minimum detachment cut required to tag the charmed
particle through its lifetime.
Fig.~\ref{fig:ptime} shows reduced proper time distributions
for the two right-sign decays
$D^0\rightarrow K^-\pi^+$ and $\bar{D}^0\rightarrow K^+\pi^-$.

\begin{figure}[t]
{\hbox
{\psfig{figure=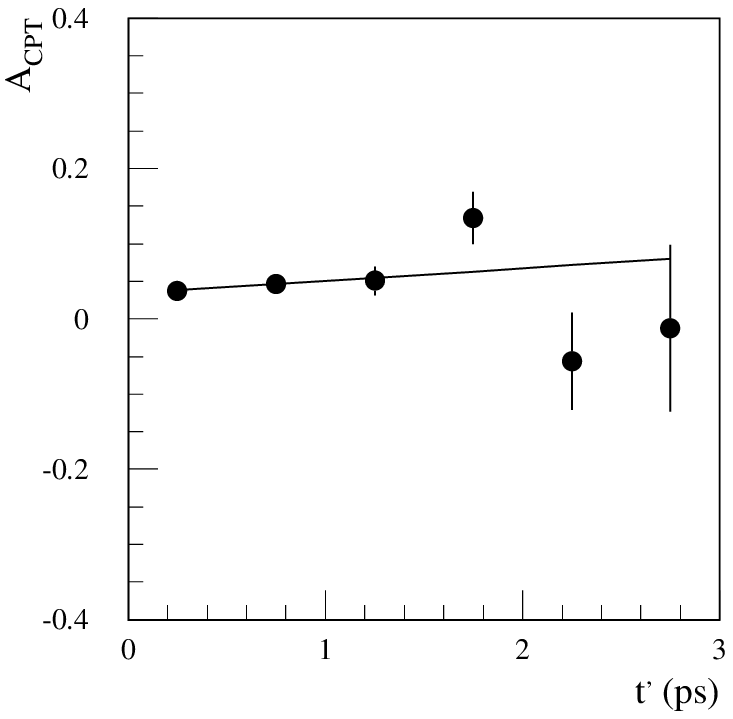,height=2.0in}
\hskip 0.25in
\psfig{figure=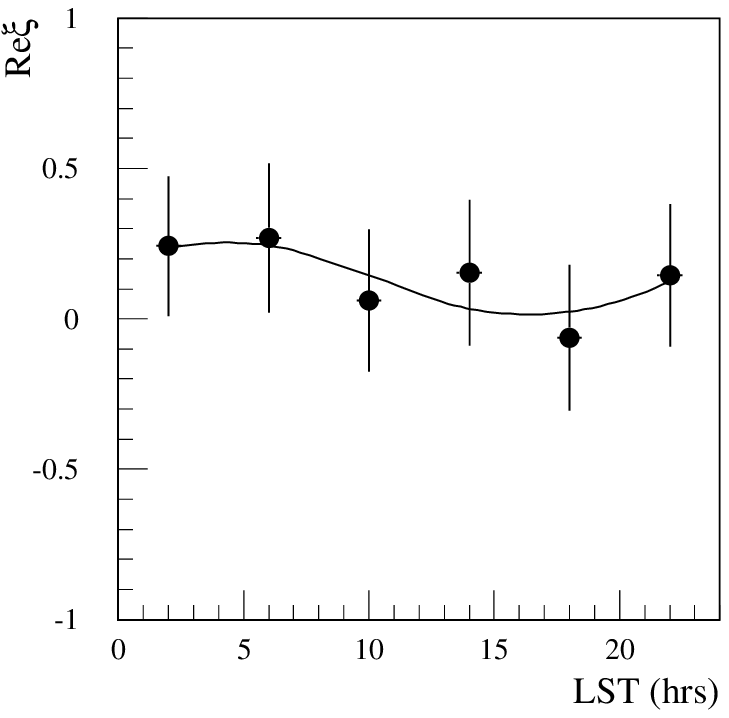,height=2.0in}}
}
\caption{Asymmetry as a function of reduced proper time (left).  At the right is
plotted the variation of the fitted CPT parameter Re~$\xi$ as a function of local
sidereal hour.}
\label{fig:asym_data}
\end{figure}

We plot the difference in right-sign events between $\d0b$ and
$D^0$ in bins of reduced proper time $t'$ in Fig.~\ref{fig:asym_data}.  
For each data point, the background subtracted yields of right-sign $D^0$ and 
$\d0b$ are used in forming the ratio. 
In the absence of detector acceptance corrections, this is 
equivalent to $A_{CPT}$ as defined in Eqn.~\ref{eq:asym}.  
Monte Carlo studies showed that detector and resolution effects cause
less than 5\% changes in the $t'$ distribution as measured by
deviations from a pure exponential decay.  $A_{CPT}$ is additionally
insensitive to detector effects since they tend to cancel in the
ratio.

Because of the QCD production mechanism for photoproduced charm mesons,  
more $\d0b$ than $D^0$ are produced in the FOCUS data sample. This has
been previously investigated in photoproduction by
E687, in which the production asymmetries were
studied in the context of a string fragmentation model.\cite{e687}  The effect
in the $A_{CPT}$ distribution is to add a constant, production-related
offset, which can be accounted for in fits to the asymmetry distributions. 

We infer the CPT sensitivity for a fixed set of experiment bounds on mixing 
without CPT violation.  The $A_{CPT}$ data in Fig.~\ref{fig:asym_data} are 
fit using the time dependent probability functions for $\d0b$ and $D^0$, 
as defined by Eq.~\ref{eq:rsdecay}
for the case of $D^0$ decay.  The allowed fit parameters are a constant
production asymmetry parameter and Re$~\xi$ (fits with Im~$\xi$ gave similar
results for sensitivity).  In the fit shown we used
values for the $x$ and $y$ mixing parameters of 5\%, and infer one
standard deviation errors on either Re~$\xi$ or Im~$\xi$ of approximately
10\%, corresponding to 95\% confidence level upper bounds of
roughly 25\%.

In the CPT and Lorentz violating extension to the  standard 
model,\cite{colladay-kostelecky}
the CPT violating parameters may depend on lab momentum, orientation, and 
sidereal time.\cite{kostelecky-prl,kostelecky}
They are expressed in terms of Lorentz violating coupling 
coefficients which depend on the flavor of the valence quark states. 
(For this reason, CPT violation in 
the $K$, $D$, and $B$ systems need not be same.)  In the case of FOCUS, 
a forward, fixed-target spectrometer, the $\xi$ parameter assumes the 
following form
\begin{eqnarray}
\xi(\hat{t}, p)  & = & {\gamma(p)\over{\Delta\lambda}}[ \Delta a_0 + \beta \Delta a_Z {\rm cos} \chi \nonumber \\
        & &
+ \beta {\rm sin} \chi (\Delta a_Y {\rm sin} \Omega \hat{t} + \Delta a_X {\rm cos} \Omega \hat{t}) ],
\label{eq:xi}
\end{eqnarray}
\noindent
where $\Omega$ and $\hat{t}$ are the sidereal frequency and time respectively, 
and $X, Y, Z$ are non-rotating coordinates with $Z$ aligned with the
Earth's rotation axis.
Our sensitivity quoted above should then be regarded as that for 
$|\overline{\xi}|$ averaged over momentum and sidereal time.  From
this quantity one can infer the
sensitivity for the Lorentz violating charm and ${\bar u}$ quark
coefficients: 
${\Delta a_0} + 0.6 \times {\Delta a_Z}  < 10^{-15}$~GeV,
where we have used 53 degrees as the angle between the FOCUS spectrometer
axis and the Earth's pole. 

We searched for a sidereal dependence, which would be sensitive to coefficients
$\Delta a_X$ and $\Delta a_Y$,
by dividing our data sample into four-hour bins in sidereal time and repeating
our fits.  The resulting distribution, shown also in Fig.~\ref{fig:asym_data},
can be fit using Eq.~\ref{eq:xi} above to yield bounds on $\Delta a_X$ and $\Delta a_Y$
of similar sensitivity.

\section{Conclusions}

We have investigated the sensitivity to a CPT violating signal in
neutral charm meson oscillations, the first search of its
kind in the charm system.
Our preliminary study using FOCUS data indicates 
95\% confidence level upper bounds on charm CPT violation 
parameters Re~$\xi$ and Im~$\xi$ of approximately 25\%, 
corresponding to bounds on the CPT and Lorentz violating 
charm coefficients of $< 10^{-15}$~GeV.

\section*{Acknowledgments}
We wish to acknowledge the assistance of the staffs of 
the Fermi National Accelerator Laboratory, the INFN of Italy, 
and the physics departments of the collaborating institutions.
The research was supported in part by the U.S. National Science
Foundation, the U.S. Department of Energy, the Italian Instituto
Nazionale di Fisica Nucleare and Ministero dell'Universita
della Ricerca Scientifica e Tecnologica, the Brazilian Conselho
Nacional de Desenvolvimento Cientfico e Tecnologico,
CONCAyT-Mexico, the Korean Ministry of Education, and
the Korean Science and Engineering Foundation.

\end{document}